\documentclass[prl,reprint]{revtex4-1}
\usepackage{dcolumn}%
\usepackage{bm}%
\usepackage{amsmath}    
\usepackage{graphicx}   
\usepackage{verbatim}   
\usepackage[usenames, dvipsnames]{color}      
\usepackage{subfigure}  
\usepackage{hyperref}   
\usepackage[normalem]{ulem}		
\begin{document} 
\title{Tunable soft-matter optofluidic waveguides assembled by light}

\author{Oto~Brzobohat\'{y}}
\email{otobrzo@isibrno.cz}
\affiliation{Institute of Scientific Instruments of the CAS. v.v.i.,Kr\'{a}lovopolsk\'{a}~147, 612~64~Brno, Czech Republic}
\author{Luk\'{a}\v{s} Chv\'{a}tal}
\affiliation{Institute of Scientific Instruments of the CAS. v.v.i.,Kr\'{a}lovopolsk\'{a}~147, 612~64~Brno, Czech Republic}
\author{Alexandr Jon\'{a}\v{s}}
\affiliation{Institute of Scientific Instruments of the CAS. v.v.i.,Kr\'{a}lovopolsk\'{a}~147, 612~64~Brno, Czech Republic}
\author{Martin \v{S}iler}
\affiliation{Institute of Scientific Instruments of the CAS. v.v.i.,Kr\'{a}lovopolsk\'{a}~147, 612~64~Brno, Czech Republic}
\author{Jan Ka\v{n}ka}
\affiliation{Institute of Scientific Instruments of the CAS. v.v.i.,Kr\'{a}lovopolsk\'{a}~147, 612~64~Brno, Czech Republic}
\author{Jan Je\v{z}ek}
\affiliation{Institute of Scientific Instruments of the CAS. v.v.i.,Kr\'{a}lovopolsk\'{a}~147, 612~64~Brno, Czech Republic}
\author{Pavel Zem\'{a}nek}
\affiliation{Institute of Scientific Instruments of the CAS. v.v.i.,Kr\'{a}lovopolsk\'{a}~147, 612~64~Brno, Czech Republic}

\date{\today}




\begin{abstract}
Development of artificial materials exhibiting unusual optical properties is one of the major strands of  current photonics research~\cite{Chen_NM_2010}. Of particular interest are soft-matter systems reconfigurable by external stimuli that play an important role in research fields ranging from physics to chemistry and life sciences~\cite{ContiPRL05, ReecePRL07,MatuszewskiOE08,FardadNL14,ManPRL13, Bezryadina_PRL_2017}.
Here, we prepare and study unconventional self-assembled colloidal optical waveguides (CWs) created from wavelength-size dielectric particles held together by long-range optical forces~\cite{DholakiaRMP10}. 
We demonstrate robust non-linear optical properties of these CWs that lead to optical transformation characteristics remarkably similar to those of gradient refractive index materials and enable reversible all-optical tuning of light propagation through the CW. 
Moreover, we characterize  strong optomechanical interactions responsible for the CW self-assembly; in particular, we report self-sustained oscillations of the whole CW structure tuned so that  the wavelength of the laser beams forming the CW is not allowed to propagate through. 
The observed significant coupling between the mechanical motion of the CW and the intensity of light transmitted through the CW can form a base for designing novel mesoscopic-scale photonic devices that are reconfigurable by light.
\end{abstract}

\maketitle

Interaction of light with a material characterized by spatially varying refractive index can alter light propagation in unusual, nonintuitive ways. 
Using the formalism of transformation optics~\cite{Chen_NM_2010}, it is possible to design artificial materials with refractive index profiles that can bend light waves almost arbitrarily; the light can even curve around a finite-size object, rendering it virtually invisible~\cite{Leonhardt_SCI_2006, Pendry_SCI_2006}.
The typical route to realizing such exotic optical metamaterials involves rather complex fabrication of metallic \cite{Cai_NP_2007} or dielectric \cite{Valentine_NM_2009} nanostructures with precise size and spatial arrangement. 
Once fabricated, such solid metamaterials cannot be reconfigured and, thus, their optical response is fixed.
On the other hand, optofluidic systems that use liquid mixtures as optical media offer widely tunable geometries and refractive index profiles adjustable through changing flow rates and/or composition of the working liquids~\cite{Zhu_LPR_2017,Yang_NC_2012}.
Flexibility of optofluidic devices can be further increased by suspending solid dielectric or metallic micro- or nanoparticles in the working liquid, which subsequently allows the local optical response of the medium to be tuned by illumination with intense light~\cite{ContiPRL05, ReecePRL07,MatuszewskiOE08,FardadNL14,ManPRL13, Bezryadina_PRL_2017}.

\begin{figure*}[htbp]
	\centering\includegraphics[width=\textwidth]{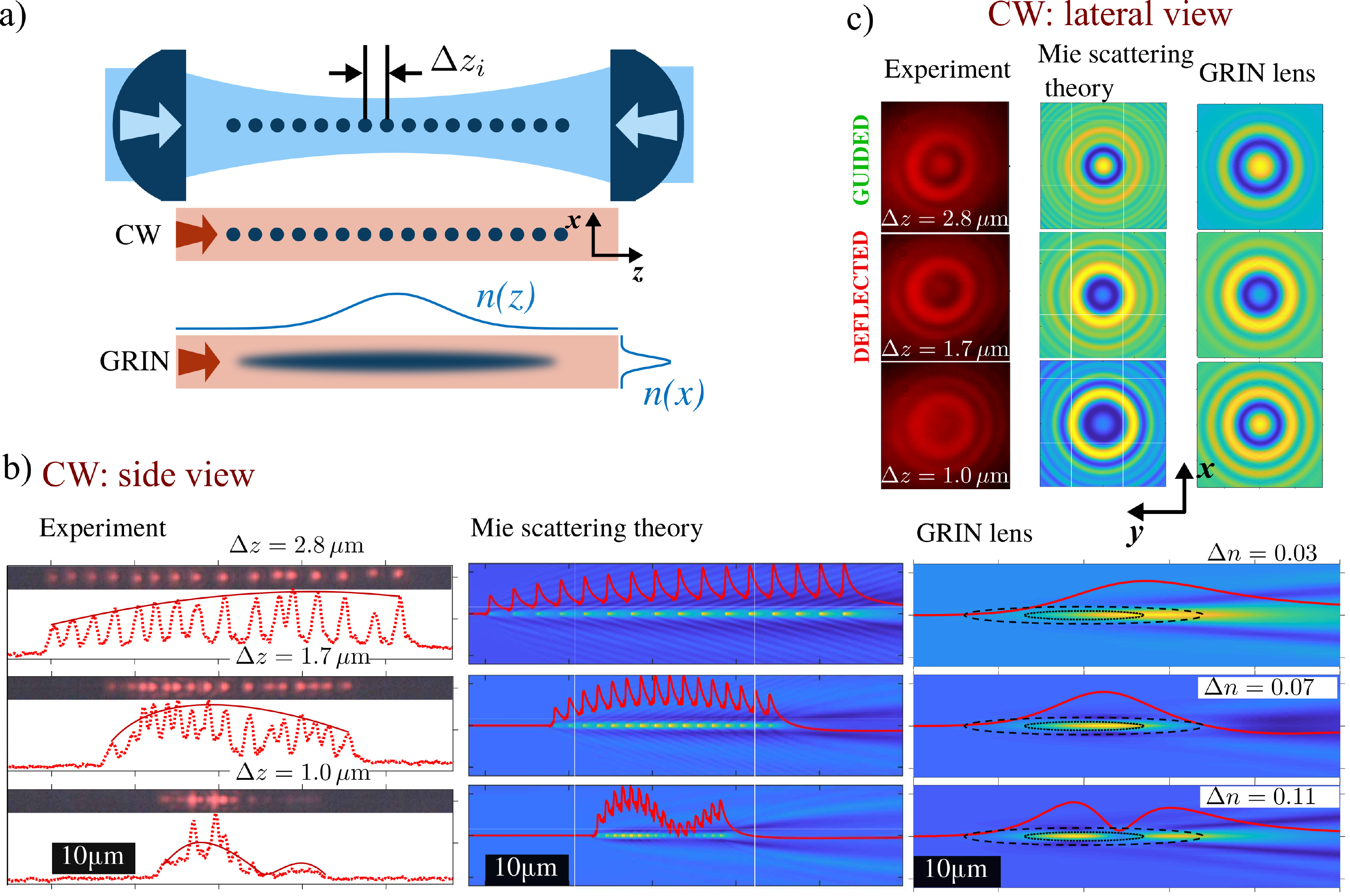}
	\caption{{\bf An optically bound CW transforms incident light similarly to a bidirectional GRIN lens} 
		a) A CW is formed by 16 identical polystyrene spheres of diameter 657\,nm suspended in water and confined in two counter-propagating trapping beams ($\lambda = 1064$\,nm; blue arrows). 
		The structure is also illuminated by an independent probe beam ($\lambda = 647$\,nm; red arrow) to examine its optical properties. 
		The discrete chain of colloidal particles can be effectively represented by a continuous GRIN lens with different profiles of the refractive index $n(z), \, n(x)$ along the axial and lateral directions, respectively.
		b) Depending on the mean inter-particle distance $\Delta z$, the red probe beam propagating through the CW is focused at the end of the structure ($\Delta z = 2.8\,\mu$m), at the front part of the structure ($\Delta z = 1.7\,\mu$m), or at two successive locations along the structure axis ($\Delta z =  1.0\,\mu$m). 
		A similar behavior is observed with bi-directional GRIN lenses featuring successively increasing axial and lateral gradients of the refractive index. 
		Smooth curves in the experimental plots indicate the envelope of the axial intensity profiles.
		c) Lateral profiles of the probe beam intensity behind the CW/GRIN lens. Depending on the position of the beam focus along the structure axis, far-field light is either concentrated on the axis (top row) or deflected into a conical region of high intensity centered on the axis (middle and bottom row). 
		\label{tuning}}
		\vspace{-0.2cm}
\end{figure*}

Here, we demonstrate soft-matter \emph{colloidal waveguides} (CWs) formed by light-induced dynamic self-assembly of micron- and submicron-sized dielectric particles immersed in a host liquid.
The internal structure of our optofluidic CWs is created by the joint action of optical gradient forces and long-range optical binding forces acting in two counter-propagating, non-uniform light beams~\cite{DholakiaRMP10,TatarkovaPRL02} [see Fig. \ref{tuning}a) for illustration of the experimental geometry].
Hence, a change in the intensity profile of the beams allows for dynamic, reversible tuning of CW optical characteristics. 
When a non-uniform light beam propagates through a colloidal suspension, the gradient forces attract high-refractive index particles toward the region of the maximal optical intensity whereas low-refractive index particles are repelled away from this region \cite{ElGanainyOE07}. 
At the same time, distances between individual particles are adjusted by the binding forces~\cite{DholakiaRMP10,KarasekPRL08,BrzobohatyAPL11}. 
Changes of the local particle concentration then result in the corresponding changes of the effective refractive index of the composite medium in an intensity-dependent fashion.
This nonlinear light-matter interaction resembling Kerr effect observed with conventional  materials can lead, for example, to self-focusing of the beam and creation of so-called optical spatial solitons which can propagate over long distances without significant diffraction~\cite{Bezryadina_PRL_2017,FardadNL14,Terborg_OL13,ManPRL13}.


\begin{figure*}[htbp]
	\centering\includegraphics[width=\textwidth]{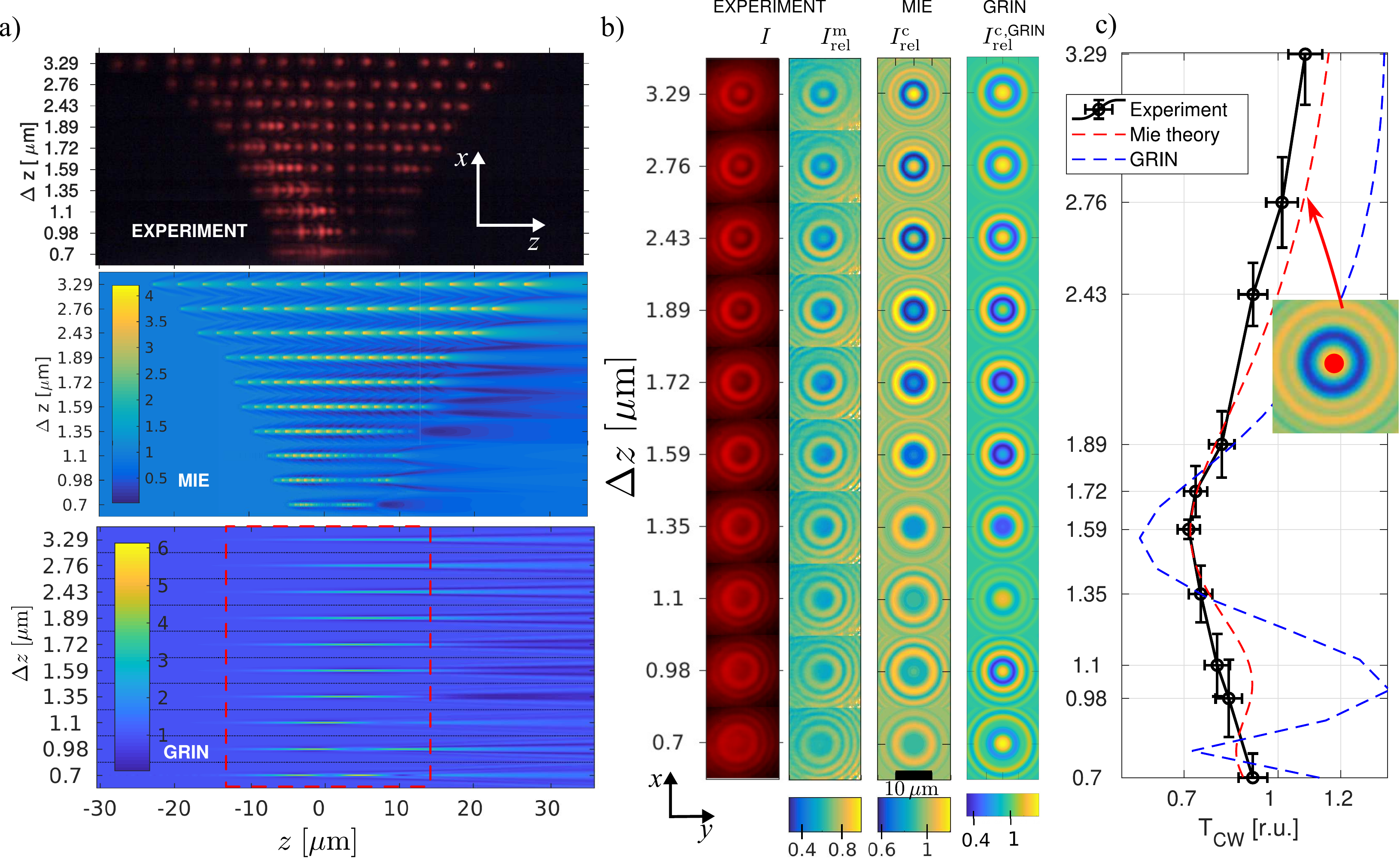}
	\caption{{\bf Transmission through optically tuned CW structures at a single probe wavelength.}   
		a) Measured (top) and simulated (middle) $xz$-plane view of CW structures with various lattice constants $\Delta z$ formed by 16 particles. 
		Experimental images were acquired at the probe-light wavelength $\lambda = 647$\,nm, using a camera with the viewing direction perpendicular to the beam propagation direction. 
		Bottom panel shows the $xz$-plane intensity distribution created by idealized GRIN lens (see text for details).
		b) (column 1) Raw measured $xy$ intensity profiles $I(x,y)$, (column 2) relative measured $xy$ intensity profiles $I^\mathrm{m}_\mathrm{rel}(x,y) = I(x,y)/I_0(x,y)$ normalized to the intensity $I_0(x,y)$ of the probe beam,
		(column 3) relative calculated $xy$ intensity profiles $I^\mathrm{c}_\mathrm{rel}(x,y)$ obtained for the same values of $\Delta z$ as in part a), (column 4) relative calculated $xy$ intensity profiles $I^\mathrm{c,GRIN}_\mathrm{rel}(x,y)$ for the GRIN lens studied in part a). 
		c) Measured (black) and calculated (red) CW transmission coefficient $T_\mathrm{CW}$ of the probe light as a function of the lattice constant $\Delta z$.  
		$T_\mathrm{CW}$ was determined by averaging the relative intensity $I^\mathrm{m}_\mathrm{rel}$, $I^\mathrm{c}_\mathrm{rel}$ of the probe light transmitted through the CW structure  over the central region of the $xy$ intensity profile denoted by a solid red circle in the inset. 
		Blue curve shows the transmission coefficient of the GRIN lens studied in part a), calculated from $I^\mathrm{c,GRIN}_\mathrm{rel}$.
		\label{tuning2}}
		\vspace{-0.2cm}
\end{figure*}

Our CWs are optically organized from individual wavelength-sized colloidal particles. Surprisingly, despite the discrete nature of such CWs, we find that their quasi-periodic structure transforms an incident light wave very similarly to a bi-directional gradient index (GRIN) lens with continuously varying refractive index profile~\cite{Gomez_Reino_2008}.
CWs formed by chains of particles with mean interparticle spacing (lattice constant) $\Delta z$ can be characterized by the effective refractive index $n \propto 1/\Delta z$. Consequently, external control of $\Delta z$ by simple adjustment of the spatial profile of the trapping beams~\cite{BrzobohatyAPL11} provides direct access to dynamic, reversible modulation of $n$ without any mechanical actuation.
This contrasts with all-liquid optofluidic GRIN structures that require control of liquid flow for changing their optical response~\cite{Zhu_LPR_2017,Yang_NC_2012}. 
Moreover, optofluidic waveguides based on colloidal suspensions are less sensitive to variations in environmental temperature and provide a wider range of accessible refractive indices determined by selecting the combination of solid particles and surrounding host liquid.

Optical back-action mediated by light scattering provides mechanical rigidity to our self-organized CWs; on the other hand, this back-action results in strong optomechanical coupling that can lead to complex dynamics of the structure and, eventually, its instability. Such optomechanical coupling has been a subject of active research in connection with developing new classes of artificial optical materials and metamaterials~\cite{RiedingerNat16,Eichenfield_NAT_2009}. In the context of optofluidic device fabrication, the dynamic behavior of CW structures can be exploited to implement novel types of optical modulators. 

To investigate the basic photonic properties of CWs formed by optical binding, we employed a weak, tunable probe beam that illuminated the CW coaxially with the two trapping beams [see Fig.~\ref{tuning}a)] and recorded the intensity profile of probe light scattered from the CW in two perpendicular directions. 
In addition, we simulated the full three-dimensional distribution of the probe-light intensity around the CW and also around an optically equivalent bi-directional GRIN lens for comparison.
The GRIN lens was defined by a three-dimensional Gaussian distribution of the refractive index: $n(x,y,z) = \Delta n  \exp[-(x^2 + y^2)/(2\sigma_r^2) -z^2/(2\sigma_z^2)]$, where $\Delta n = C (n_\mathrm{sph}-n_\mathrm{med}) a/\Delta z$, $a$ is the particle radius, $n_\mathrm{sph} = 1.59$ and $n_\mathrm{med} = 1.33$ are the refractive indices of the particle and the surrounding medium, respectively, $\sigma_r = \sigma_x = \sigma_y$ and $\sigma_z$ are the widths of the distribution in the $x(y)$\,- and $z$\,-directions, and $C$ is a proportionality constant.
In the GRIN lens simulations, we set $\sigma_z = 7\,\mu$m, $\sigma_r = 0.5\,\mu$m, and $C=1.05$ to obtain the best agreement with the CW experiments (See Supplementary information for more details).


\begin{figure*}[htbp]
	\centering\includegraphics[width=0.95\textwidth]{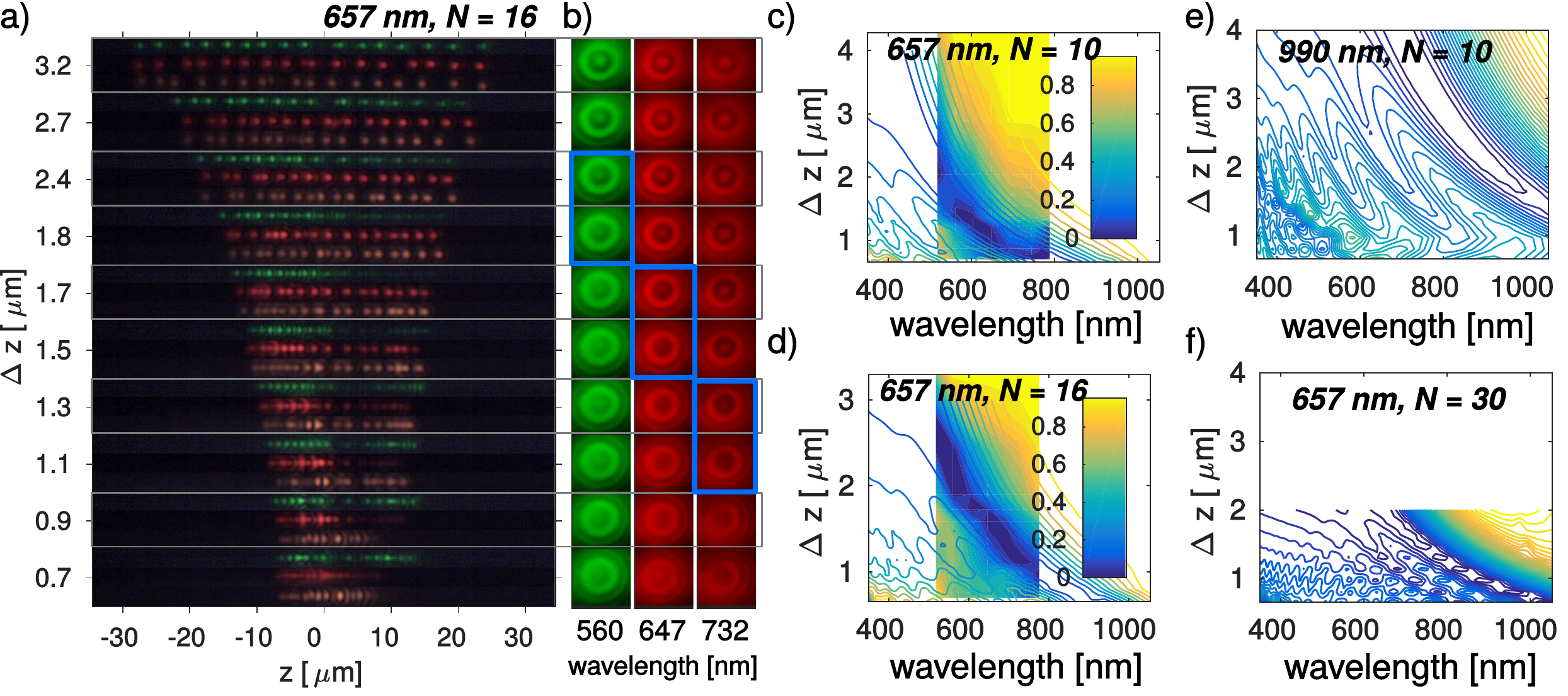}
	\caption{{\bf Tuning of  light transmission characteristics of CWs by changing  inter-particle distance,  number of particles, and size of particles.} a) $xz$ and b) $xy$ views of intensity profiles of the probe light with three different wavelengths (560 nm, 647 nm, and 732 nm) scattered from a CW structure formed by $N=16$ polystyrene spheres with the diameter of 657 nm. 
	Blue rectangles in part b) highlight CW configurations with the minimal transmission coefficient $T_\mathrm{CW}$ at the studied wavelength corresponding to the deflecting transmission modes.	
	c) and d) Comparison of experimentally (colored background) and theoretically (colored contours) determined spectral profiles of $T_\mathrm{CW}$ for a wide range of inter-particle distances $\Delta z$ in CWs formed by 10 spheres [part c)] and 16 spheres [part d)] of diameter 657 nm. For each wavelength, the values of $T_\mathrm{CW}$ have been rescaled to the range $\left\langle 0;1\right\rangle$ (see Supplementary information). 
	e) and f) Theoretically determined spectral profiles of the rescaled $T_\mathrm{CW}$ for CWs formed by 10 spheres of diameter 990 nm [part e)] and 30 spheres of diameter 657 nm [part f)]. 
		\label{bandgap}}
	\vspace{-0.2cm}
\end{figure*}

To characterize the transmission of the probe light through the CW structure, we introduce CW guiding and deflecting modes.
The light is considered to be guided when the maximum of its far-field intensity in the lateral direction (the $xy$-plane) lies on the CW axis (the $z$-axis); conversely, deflected light is characterized by an off-axis intensity maximum. 
Due to the axial symmetry of CW structures, the deflected light forms a hollow conical shell centered on the CW axis and, thus, its $xy$ intensity profile has the form of a bright ring.
Figures~\ref{tuning}b) and~\ref{tuning}c) illustrate the change of the transmitted light pattern when the value of $\Delta z$ is reduced. 
Upon reducing $\Delta z$, the effective refractive index of the CW increases and the focus of the probe beam ($\lambda = 647$\,nm) moves from the end of the structure ($\Delta z = 2.8\,\mu$m) to the front part of the structure ($\Delta z = 1.7\,\mu$m), eventually creating two successive foci along the CW axis ($\Delta z =  1.0\,\mu$m).
This motion of the axial position of the probe beam focus is accompanied by change of the transmitted light pattern from guided to deflected.

As shown in the middle column of  Figs.~\ref{tuning}b) and \ref{tuning}c), calculated  profiles of the probe light intensity lead to the same qualitative conclusions as experimental data.
Remarkably, the comparison of CW to bi-directional GRIN lens reveals very similar optical transformation properties of the two structures, even though the spatial profile of refractive index of the GRIN lens has an idealized form of a continuous two-dimensional Gaussian distribution with constant widths $\sigma_r$ and $\sigma_z$ [see right column of Figs.~\ref{tuning}b) and \ref{tuning}c)].

Figure~\ref{tuning2} summarizes a parametric experimental and theoretical study of interaction of fixed-wavelength probe light with CW structures carried out for different values of the CW lattice constant $\Delta z$ and, consequently, different effective refractive index profiles. 
In order to quantify the transmission properties of CWs with different lattice constants at a fixed probe wavelength, we calculated the transmission coefficient $T_\mathrm{CW}$ of CW structures at the probe wavelength, using the procedure described in the caption of Fig.~\ref{tuning2}. 
In terms of $T_\mathrm{CW}$, the deflecting case corresponds to the minimal value of $T_\mathrm{CW}$ while the guiding case is characterized by the maximal value of $T_\mathrm{CW}$.

Figure~\ref{tuning2}c) shows the variations of the measured and calculated $T_\mathrm{CW}$ with $\Delta z$.
We obtained an encouragingly good agreement between the experimental and theoretical values of $T_\mathrm{CW}$, even though the calculations assumed equal inter-particle distances in the CW which are typically not observed in the experiments [see Fig.~\ref{tuning2}a) for illustration]. 
In general, non-uniformity of inter-particle spacing increases with closer particle packing; this trend leads to a larger difference between the measured and simulated $T_\mathrm{CW}$ for smaller $\Delta z$ (see Supplementary information).
In addition, our simulations again show that the CW transmission properties are  similar to those of bi-directional GRIN lenses. 

In order to directly visualize the dependence of the spectral position of guiding and deflecting CW transmission modes on $\Delta z$, we recorded the intensity of the probe light transmitted through CWs with different $\Delta z$ at several wavelengths in the visible spectral region. 
The comparison of intensity profiles recorded at different probe wavelengths [Figs.~\ref{bandgap}a) - \ref{bandgap}b)] reveals the dispersion of the effective refractive index $n$ of the CW structures: as indicated by the blue rectangles in Fig. ~\ref{bandgap}b), with increasing wavelength, the deflecting mode shifts towards smaller values of $\Delta z$.

A detailed experimental and theoretical study of the dependence of CW transmission characteristics on the inter-particle distance, the number of particles, and the size of particles is presented in Figs.~\ref{bandgap}c) - \ref{bandgap}f).
\begin{figure*}[htb]
	\centering\includegraphics[width=0.7\textwidth]{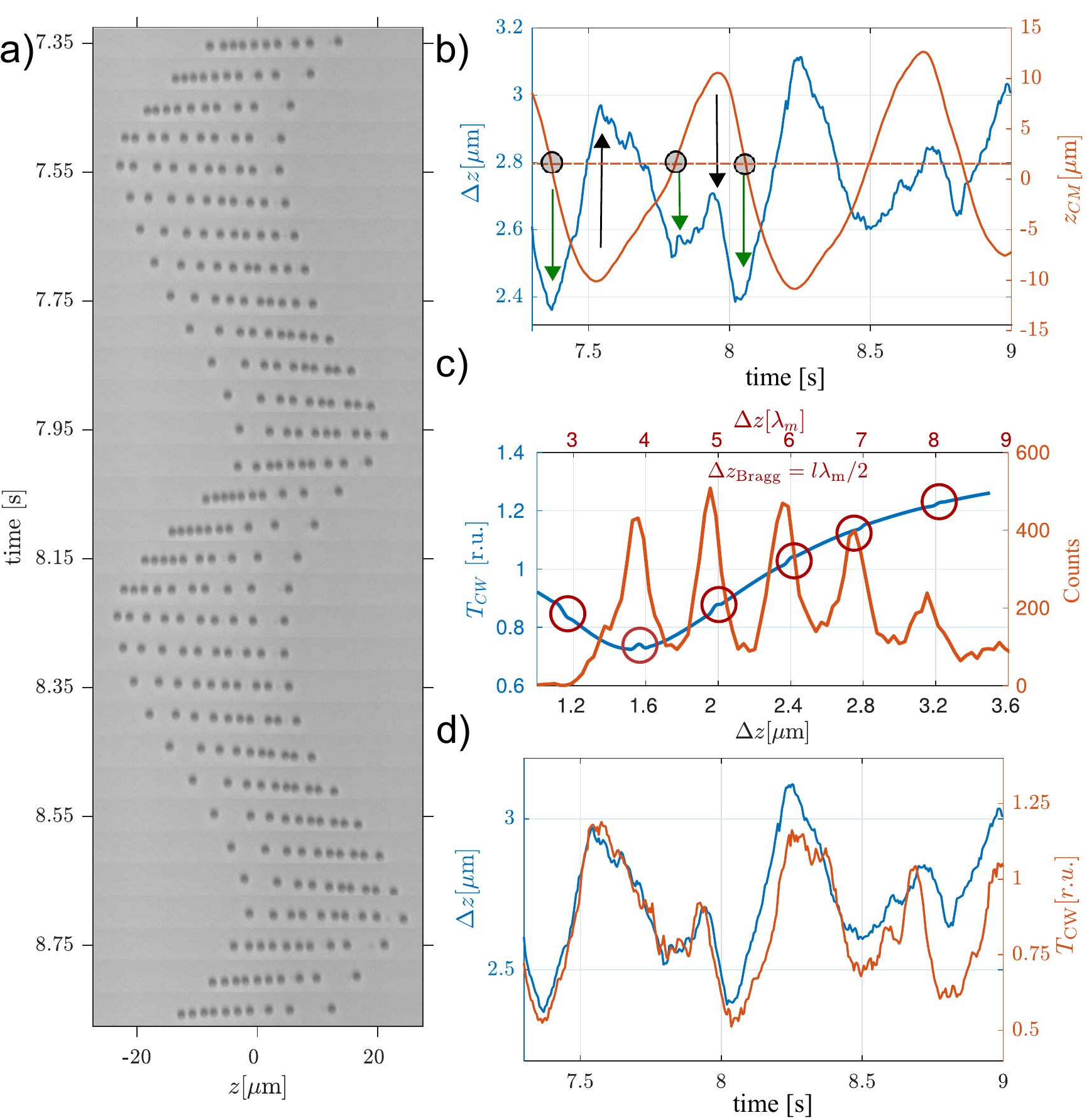}
	\caption{{\bf Self-sustained oscillations of a CW structure deflecting the trapping light.} a) Time sequence of images of an oscillating CW formed by 10 particles (diameter 990 nm), with deflecting mode at the trapping light wavelength of 1064\,nm. See also Media3. 
	b) Mean inter-particle distance $\Delta z$ (blue curve) and position of the CW center of mass $z_\mathrm{CM}$ (red curve) observed for the oscillating CW shown in part a).
	The CW structure breathes at twice the oscillation frequency of its center of mass. 
	Successive extremes of $\Delta z$ are highlighted by arrows, zeros of $z_\mathrm{CM}$ are highlighted by filled circles. 
	c) [blue] Transmission coefficient $T_\mathrm{CW}$ calculated for the trapping wavelength of 1064\,nm and uniform inter-particle distance $\Delta z$ along a CW formed by 10 particles with the diameter of 990\,nm. 
	Red circles indicate the values of $\Delta z$ that fulfill Bragg condition: $\Delta z = \Delta z_{\mathrm{Bragg}} = l\lambda_\mathrm{m}/2$. 
	[orange] Histogram of inter-particle distances $\Delta z$ observed in the experiments during self-sustained CW oscillations. 
	On average, inter-particle spacing in the oscillating structure fulfills Bragg condition.
	d) Dynamics of CW transmission at the trapping light wavelength $T_\mathrm{CW}$ (red curve) is directly coupled to the breathing of the CW structure $\Delta z$ (blue curve).
	\label{optomech}}
	\vspace{-0.2cm}
\end{figure*} 
In general, for a constant inter-particle spacing $\Delta z$, an increase of the CW length due to increasing number of the particles [compare parts c), d), and f)] or an increase of the effective refractive index of the CW due to increasing diameter of the particles [compare parts c) and e)] causes a shift of the spectral position of the deflecting mode toward longer wavelengths.
When the number and size of the particles are kept constant, as typical in an experiment, the spectral position of the deflecting mode can be shifted toward longer wavelengths by increasing the effective refractive index due to decreasing $\Delta z$, which can be accomplished by reducing the radius of the trapping beams \cite{BrzobohatyAPL11}. 
All these characteristics can be replicated by modelling the CW as an equivalent bi-directional GRIN lens with suitably chosen profiles of the refractive index (see Supplementary information).

The change of the CW transmission characteristics from guiding to deflecting has an important implication for the stability of the CW structures formed by optical binding. 
When the distance between the particles forming the CW  is gradually reduced (i.e. the effective refractive index $n$ is increased), spectral position of the deflecting mode eventually approaches the trapping wavelength at 1064\,nm. 
Since the trapping light now cannot propagate through the CW structure and is instead deflected off the CW axis, the confinement of the particles caused by optical trapping and binding forces is momentarily lost and the CW structure briefly falls apart due to Brownian motion. 
However, as soon as the structure starts losing its form, its effective refractive index changes and the trapping beam can again propagate through the structure, thus restoring the original order. 
The whole cycle can be then repeated over and over. 
This optomechanical feedback, which strongly couples the arrangement of the particles in the CW structure with its photonic properties and stability, turns CWs into a highly dynamic oscillatory system with complex phase-space trajectories. 

In order to study experimentally the onset of optomechanically induced instability of CW structures, we followed the theoretical predictions presented in Fig.~\ref{bandgap} and approached deflecting mode at the trapping wavelength by increasing the number of particles in the CW while keeping the trapping beam width fixed. 
Under these experimental conditions, spectral position of deflecting mode shifts to a longer wavelength due to the simultaneously increasing length and effective refractive index of the waveguide.
Indeed, once the number of particles in the chain reaches a critical value ($N\approx40$ for the given particle size and radius of the trapping beams) for which the trapping beams are deflected, the CW structure becomes unstable and starts oscillating~\cite{GordonPRB08,TatarkovaPRL02} (see Media1 and Media2 for illustration of the oscillation dynamics). 

Systematic experimental investigation of optomechanically induced instability in CWs formed by large numbers of particles is rather challenging.
Thus, guided by the theoretical prediction of Fig.~\ref{bandgap}e), we took an alternative route and reached the deflecting mode at the trapping wavelength by forming a CW from 10 larger particles  (diameter 990 nm) and gradually decreasing the mean inter-particle distance $\Delta z$ by reducing the width of the trapping laser beams.

Figure~\ref{optomech}a) reveals that the particle spacing in such an unstable CW is highly variable; the particles are non-uniformly squeezed together and the axis of the formed CW is slightly tilted with respect to the common optical axis of the trapping beams. Furthermore, the structure displays oscillatory behaviour and both the position of its center of mass and the configuration of individual particles within the structure periodically change (see also Media3).

As illustrated by Fig.~\ref{optomech}b), the frequency of periodic deformations of the CW structure described by the value of mean inter-particle distance $\Delta z$ (blue curve) is two times higher than the frequency of oscillations of the CW center of mass $z_\mathrm{CM}$ (red curve). 
This characteristic oscillation pattern is consistent with the spatial symmetry of two counter-propagating trapping beams supplying external optical forces.
As argued by Taylor and Love~\cite{Taylor_PRA_09}, spontaneous symmetry breaking leading to periodic circulation of optically bound structures can occur in otherwise symmetric counter-propagating beam geometry for particles with a sufficiently large size and/or relative refractive index. 
In our case, this breaking of symmetry in optical forcing is further amplified by the nonlinear optical response of the CW to the guided light.
The above described behaviour represents a manifestation of the strong optomechanical feedback that destabilizes the CW structure once its deflecting mode approaches the trapping light wavelength, as indicated in Fig. \ref{optomech}c).
The histogram of experimental inter-particle distances $\Delta z$ observed in an oscillating CW structure [orange curve in Fig.~\ref{optomech}c)] clearly shows that during oscillations, particles are optically bound preferentially at separations $\Delta z = l\lambda_\mathrm{m}/2$~\cite{KarasekPRL08}. 
Here, $l$ is a whole number and the trapping light wavelength $\lambda_\mathrm{m}$ in the effective medium characterizing the CW is reduced with respect to the vacuum wavelength by the refractive index of the CW medium.
This preferred inter-particle spacing corresponds to Bragg condition: $\Delta z = \Delta z_{\mathrm{Bragg}} = l\lambda_\mathrm{m}/2$. 
Thus, the quasi-periodic CW structure effectively acts as a weak photonic crystal (PC), enhancing the scattering of light with wavelength commensurate with the lattice constant of the structure \cite{Joannopoulos11}. 
Indeed, the simulated CW transmission curve shown in Fig.~\ref{optomech}c) reveals  small local maxima of $T_\mathrm{CW}$ superimposed on a slowly varying background that arises from the changes of the effective refractive index. 
These discrete transmission peaks are separated by $\Delta z_{\mathrm{Bragg}}$ and originate from constructive interference of light scattered from the CW structure.  
However, with the number and relative refractive index of particles used in our experiments and simulations, the PC effect is rather weak and contributes only marginally to the overall transmission characteristics of our CW structures.

During the self-sustained oscillation cycle, the CW alternately deflects and guides the trapping beams. Figure~\ref{optomech}d) shows the precise correlation between the measured transmission coefficient $T_\mathrm{CW}$ of the trapping light propagating through the structure (red curve) and the periodic deformations of the CW structure that cause corresponding modulation of the CW refractive index (blue curve). 
This optomechanical coupling between the intensity of light transmitted through the CW and the motion of the CW can serve as a control mechanism in integrated photonic devices that are fully actuated by light.
Among the potential applications of this feedback/modulation scheme, one can envision spectral-mode filtering in hollow-core photonic fibers using CWs self-arranged in the fiber core \cite{Benabid_OE_02}.





In summary, we have introduced colloidal waveguides that self-arrange from wavelength-sized dielectric particles suspended in an aqueous environment due to the exchange of momentum between the particles and light fields of counter-propagating trapping laser beams.
By adjusting the properties of the trapping beams, the transmission spectrum of these CWs can be reversibly tuned. 
Upon configuring the CWs so that their deflecting transmission mode approaches the trapping wavelength, internal organization of the waveguides changes from static to dynamic, which has a profound impact on their optical characteristics and can serve for implementing novel applications of these photonic structures.
Even though the proposed CWs are discrete, with quasi-periodic spatial structure formed by individual colloidal particles, their light-transformation characteristics are quite similar to continuous bi-directional GRIN lenses. 
This similarity can facilitate theoretical description of photonic elements based on discrete particle chains, thus contributing to optimization of these elements for a specific task.

While we studied the photonic and optomechanical properties of our colloidal waveguides in overdamped dynamic regime (constituent particles immersed in water), we believe there are no fundamental obstacles that would prevent the transfer of the system to vacuum. 
Thus, our photonic structures can be combined with fundamental studies of optomechanical properties of high-quality mechanical oscillators represented by optically levitated nanoparticles \cite{GieselerNatPhys13,LiNP11}.

Formation of CWs from non-spherical or chiral particles could straightforwardly extend the optical self-assembly approach towards one- or higher-dimensional birefringent or chiral non-linear systems for light transformation with yet richer spectrum of optomechanical behaviour.




\section*{Methods}

\paragraph{Simulations of CW transmissivity.}

Distributions of optical intensity due to light scattering from CW structures presented in Figs.~\ref{tuning} and \ref{tuning2} were calculated using multiple-spheres Mie theory \cite{Mackowski2012} with the help of the localized approximation for a paraxial Gaussian beam \cite{Gouesbetbook}.

2-D transmission spectra of CW structures presented in Fig.~\ref{bandgap} were calculated numerically using the finite-difference time-domain (FDTD) method \cite{taflove2000computational} implemented in a freely available software package Meep\cite{OskooiRo10}.
In order to simplify the calculations, we approximated the inhomogeneous incident beam with a Gaussian transversal intensity profile by a uniform plane wave.
The most efficient way of computing the spectral response of CW structures over a broad wavelength range is by Fourier-transforming the response of the structure to a short light pulse. 
We employed an incident pulse with a Gaussian envelope in the time domain (Gaussian width: 1.75 fs, central wavelength 695\,nm) to obtain CW transmission spectra over the spectral region between 340 and 1050\,nm. 
In the calculations, inter-particle distance $\Delta z$ was varied in 50 nm steps.

In order to calculate the transmission coefficient $T_\mathrm{CW}$ of the CW structures for the probe beam, we averaged the relative intensity of the probe beam transmitted through the CW structure [see Fig.~\ref{tuning2}~b) for illustration] over the central region of the $xy$ intensity profile denoted by a red circle in the inset of Fig.~\ref{tuning2}c).

GRIN lens calculations were performed using commercial Finite Element Method software Comsol Multiphysics. 
Light propagation was modeled by the beam envelopes method that solves the Helmholtz equation using the assumption of a slowly spatially varying amplitude (envelope) of the propagating light that is modulated by fast phase oscillations along the beam propagation axis, i.e. the electric field intensity is represented as 
\begin{equation}
  E(x,y,z) =\psi_0(x,y,z)\exp(-ikz),
\end{equation} 
where $\psi_0$ is the slowly varying beam envelope and $k$ is the wave number in the immersion medium surrounding the GRIN lens.
The refractive index is modeled as 3D Gaussian
\begin{equation}
 n(x, y, z) = n_0 + \Delta n \exp\left( - \frac{(x^2 -y^2)}{2\sigma_r^2}  - \frac{z^2}
{2\sigma_z^2} \right),
\end{equation} 
where $\Delta n$ is maximal refractive index change (see main text) and $
\sigma_r = \sigma_x =  \sigma_y$, $\sigma_z$ are profile half-widths.
The solutions of the beam envelope method were studied on a quarter-cylinder like domain (cylinder radius 75 $\mu$m, height 90 $\mu$m) where we employed symmetry of the problem with respect to polarization. 
The computational domain was filed with prism mesh swept from the entry face and was enclosed by the perfectly matched layers.
%
The Gaussian beam that would propagate through the undisturbed medium was 
excited from the input port of the computational domain.
We performed a parametric study for a given combinations of incident field 
wavelengths $\lambda$ and refractive index change $\Delta n$  (give by inter-
particle distance $\Delta z$.

\begin{figure}[h]
	\centering
	\includegraphics[width = 0.5\textwidth]{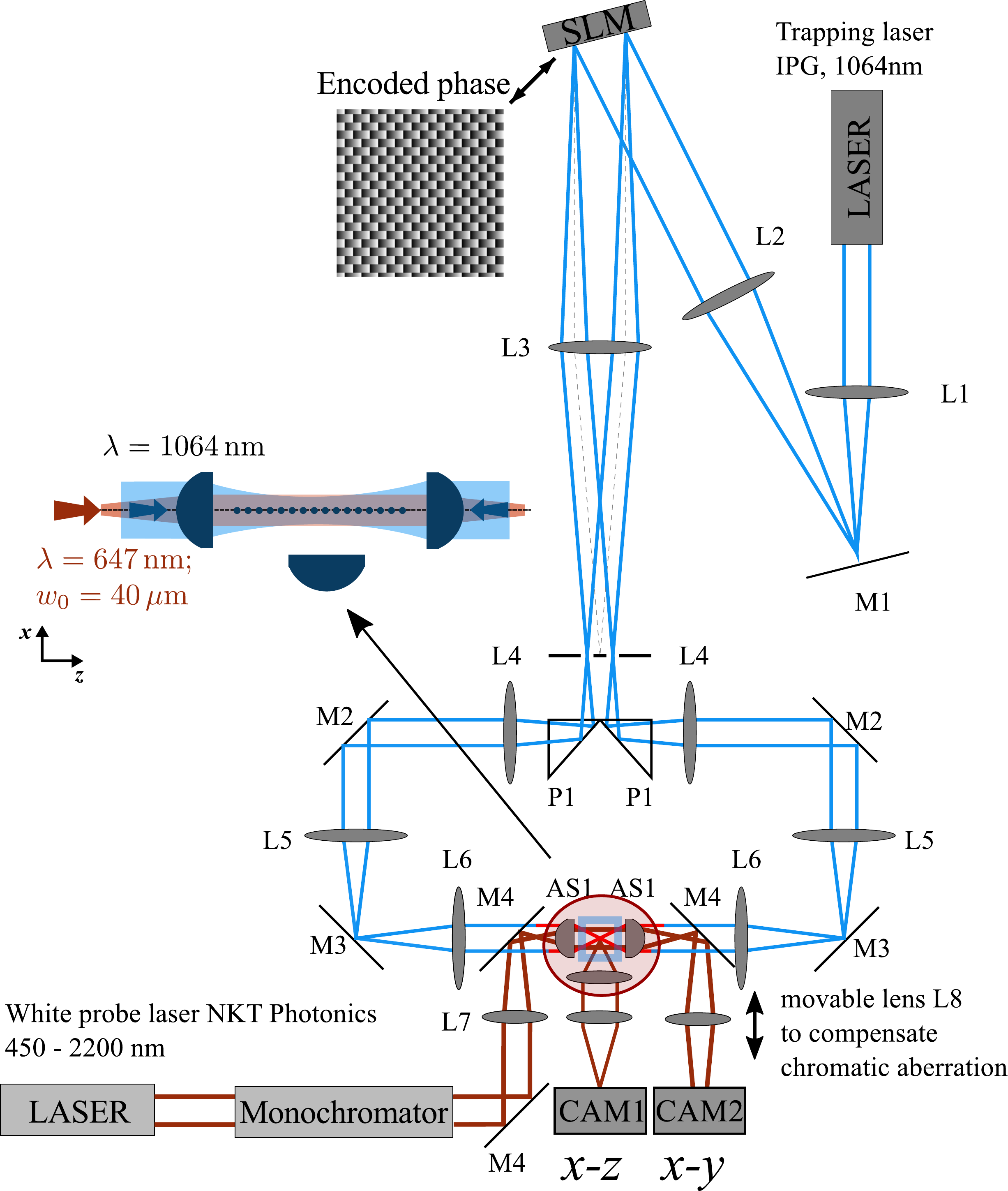}
	\caption{{\bf Experimental setup for the characterization of optically tunable CWs, consisting of the trapping (blue) and probe (red) beam paths.} 
	} 
	\label{setup}
\end{figure}

\paragraph{Experimental setup.} The setup used in our experiments with optically tunable CWs is shown in Figure~\ref{setup}.

\medskip

\noindent {\it Light path for optical trapping} \\
We used Thorlabs achromatic doublets with antireflection coating  ACN254-XXX-C (L1 -- L6), dielectric mirrors PF10-03 (M1 -- M3) and aspheric lenses C240TME-C with antireflection coating (AS1).
A collimated Gaussian beam from an infrared laser (IPG ILM-10-1070-LP; wavelength 1064 nm, maximal output power 10 W) was expanded by a telescope formed by lenses L1 ($f_1= 150$~mm) and L2 ($f_2=300$~mm) and projected on a spatial light modulator (SLM) (Hamamatsu LCOS X10468-07).
Phase mask encoded at the SLM diffracted the beam into the $\pm1$ diffraction orders that were used to generate the two counter-propagating trapping beams; the zeroth and higher orders were blocked by a stop placed in the focal plane of lens L3 ($f_3=400$~mm).
The two transmitted $1^{st}$--order beams were reflected from prisms P1 and collimated by lenses L4 ($f_4=200$~mm). These lenses formed  telescopes with the lens L3, projecting the SLM plane on the mirrors M2. 
The SLM plane was then imaged onto the back focal planes of aspheric lenses AS1 ($f=8$~mm, maximal NA = 0.5) by telescopes consisting
of lenses L5  ($f_5=100$~mm) and L6  ($f_6=150$~mm).  
Finally, AS1 focused the two trapping  beams into a square glass capillary (blue square in Fig. \ref{setup}) containing an aqueous dispersion of polystyrene spheres. 
Widths of the focused trapping beams in the sample chamber could be controlled by adjusting the area of the diffraction grating imposed upon the SLM.

\smallskip
\noindent {\it Light path for the probe beam} \\
A collimated beam from a white-continuum laser (NKT Photonics, SuperK EXTREME series; wavelength range 450-2200 nm) spectrally filtered by a monochromator (CVI, 1/2m Digikr\"{o}m) was used as a source of probe light with the wavelength tunable in the visible and near--infrared spectral region.  
The beam exiting from the monochromator was focused to the back focal plane of aspheric lens AS1 (using lens L7 with $f_7= 250$~mm and dichroic mirror M4) to produce a relatively wide probe beam with the radius $\sim$40\,$\mu$m in the trapping region.
This weak probe beam co-propagated with the trapping beams and interacted with the CW structure created in the trapping region. 
As the typical intensity of the probe beam at the CW location was approximately three orders of magnitude lower than the intensity of the trapping beams, the influence of the probe beam on the configuration of particles in the CW could be safely neglected. 
The probe light scattered from CW was imaged from two perpendicular directions using two color cameras (CAM1 and CAM2; Basler acA2000-50gc) while the scattered trapping light was blocked in the imaging optical path using a notch filter. 
In particular, the  scattered light in the $xz$--plane containing the CW axis was observed using a microscope composed of a long--distance objective (Mitutoyo M Plan Apo SL 80X), tube lens (focal length 200 mm), and CAM1.
The distribution of light in the $xy$--plane perpendicular to the CW axis was then observed employing one of the aspheric lenses AS1 as the microscope objective, Thorlabs achromatic doublet ACN254-200-A as the tube lens L8, and CAM2.
Due to chromatic aberrations of our optical system, the position of the imaging focal plane for the probe light shifted along the optical axis when the probe wavelength was changed.
To compensate for this effect, for each probe wavelength, we adjusted the position of the tube lens L8 and the camera CAM2 so as to record the scattering pattern of the probe light from the same plane behind the CW structure.

\paragraph{Sample preparation.}
Polystyrene spheres (Duke Scientific, diameters 657\,nm, 520\,nm or 990\,nm) were diluted in distilled water and sonicated for several minutes in an ultrasonic bath. 
Subsequently, the samples were loaded into a vertically mounted glass capillary with a square cross-section and the inner diameter of 100\,$\mu$m (VitroTubes; VitroCom) serving as the experimental sample chamber.


\noindent{\bf \large Acknowledgement}\\
The research was supported by
projects of CSF (GA18-27546S), and its infrastructure
by MEYS CR, EC, and CAS (LO1212, CZ.1.05/2.1.00/01.0017, RVO:68081731).


%
%


\end{document}